\documentclass[12pt,twocolumn]{openjournal}

\usepackage{xcolor}
\usepackage{textgreek}
\usepackage[utf8]{inputenc}
\usepackage[english]{babel}
\usepackage{hyperref}
\hypersetup{
    colorlinks=true,
    linkcolor=blue,
    filecolor=blue,      
    urlcolor=blue,
    citecolor=blue,
}
\usepackage{color,colortbl}
\usepackage{tensind}
\tensordelimiter{?}
\DeclareGraphicsExtensions{.bmp,.png,.jpg,.pdf}
\usepackage{verbatim}
\usepackage[normalem]{ulem}
\usepackage{orcidlink}
\usepackage{soul}

\graphicspath{ {./figs/} }

\begin{document}
\title{LIGHTS. The extended point spread functions of the LIGHTS survey at the LBT}

\shorttitle{The extended PSFs of the LIGHTS survey}
\shortauthors{Sedighi et al.}

\author{Nafise Sedighi\,\orcidlink{0009-0001-9574-8585}$^{1,2}$}
    \author{Zahra Sharbaf\,\orcidlink{0009-0004-5054-5946}$^{1, 2}$} 
    \author{Ignacio Trujillo\,\orcidlink{0000-0001-8647-2874}$^{1, 2}$} 
    \author{Sepideh Eskandarlou\,\orcidlink{0000-0002-6672-1199}$^{3}$} 
    \author{Giulia Golini\,\orcidlink{0009-0001-2377-272X}$^{1, 2}$} %
    \author{Ra\'ul Infante-Sainz\,\orcidlink{0000-0002-6220-7133}$^{3}$} %
    \author{Samane Raji\,\orcidlink{0000-0001-9000-5507}$^{4}$}
    \author{Dennis Zaritsky\,\orcidlink{0000-0002-5177-727X}$^{5}$}
     \author{Pedram Ashofteh Ardakani\,\orcidlink{0000-0002-3790-5837}$^{6}$} 
    \author{Nushkia Chamba\,\orcidlink{0000-0002-1598-5995}$^{7}$} 
   \author{S. Zahra Hosseini-ShahiSavandi,\orcidlink{0000-0003-3449-2288}$^{8}$} 
    \author{Richard Donnerstein\,\orcidlink{0000-0001-7618-8212}$^{5}$}
    \author{Mauro D'Onofrio\,\orcidlink{0000-0001-6441-9044}$^{8}$}
    \author{Garreth Martin\,\orcidlink{0000-0003-2939-8668}$^{9}$}
    \author{Mireia Montes\,\orcidlink{0000-0001-7847-0393}$^{10}$}
    \author{Javier Rom\'an\,\orcidlink{0000-0002-3849-3467}$^{11}$}

\affiliation{$^{1}$Instituto de Astrof\'isica de Canarias, C/ V\'ia L\'iactea,38200 La Laguna, Tenerife, Spain}
\affiliation{$^{2}$ Departamento de Astrof\'isica, Universidad de La Laguna, 38206 La Laguna, Tenerife, Spain}
\affiliation{$^{3}$Centro de Estudios de F\'isica del Cosmos de Arag\'on (CEFCA), Plaza San Juan, 1, E-44001, Teruel, Spain}
\affiliation{$^{4}$ Departamento de F\'sica Te\'orica, At\'omica y \'Optica, Universidad de Valladolid, 47002 Valladolid, Spain}
\affiliation{$^{5}$Steward Observatory, University of Arizona, 933 North Cherry Avenue, Tucson, AZ 85721-0065, USA}
\affiliation{$^{6}$ School of Astronomy, Institute for Research in Fundamental Sciences (IPM), Tehran, Iran}
\affiliation{$^{7}$ NASA Ames Research Center, Moffett Field, CA 94035, USA}
\affiliation{$^{8}$ Department of Physics and Astronomy, University of Padova, Vicolo Osservatorio 3, I-35122, Italy}
\affiliation{$^{9}$ School of Physics and Astronomy, University of Nottingham, University Park, Nottingham NG7 2RD, UK}
\affiliation{$^{10}$ Institute of Space Sciences (ICE, CSIC), Campus UAB, Carrer de Can Magrans, s/n, 08193 Barcelona, Spain}
\affiliation{$^{11}$ Departamento de Física de la Tierra y Astrofísica, Universidad Complutense de Madrid, E-28040 Madrid, Spain}

\begin{abstract}
    With the arrival of the next generation of ultra-deep optical imaging surveys reaching $\mu_V$$\sim$30 mag/arcsec$^2$ (3$\sigma$; 10\arcsec$\times$10\arcsec), the removal of scattered light due to the point spread function (PSF) effect remains a critical step for the scientific exploitation of the low surface brightness information contained in these data. Because virtually all pixels in the ground-based images are affected by an unwanted screen of light with a brightness greater than $\mu_V$$\sim$29 mag/arcsec$^2$, the characterization of the extended PSF (R$>$5 arcmin)  is mandatory. We describe the procedure used to construct the extended PSFs of the LIGHTS survey in the g- and r-band images taken with the Large Binocular Cameras (LBCs) of the Large Binocular Telescope (LBT). We produce PSFs with a radial extension of 6.5 arcmins. These are later extended to 30 arcmins following an empirically motivated power-law extrapolation of their behaviour in their outermost regions. As an example of the application of our methodology, we subtract the scattered light around the galaxy NGC3198. The result of this subtraction clearly shows the outermost parts of the galaxy's disc, which have been obscured by the influence of nearby bright stars.
We make all the PSF (compact and extended) models publicly available.
\end{abstract}


\maketitle

\section{Introduction}
\label{sec:intro}

During this decade, an increasing number of ultra-deep optical imaging surveys will be able to observe a large part of the night sky down to a surface brightness of $\mu_V$$\sim$30 mag/arcsec$^2$ or even fainter \citep[see e.g. Euclid, LSST;][]{Euclid2024,2024MNRAS.528.4289W,2022AAS...24030406B,2018arXiv181204897L}. The scientific use of these new images in the very low surface brightness regime ($\mu_V$$>$28 mag/arcsec$^2$) will depend on our ability to remove scattered light from the brightest sources in the images. As deeper and deeper images of some regions of the sky have become available \citep[see e.g.][]{2012ApJS..200....4F,2013ApJ...762...82M,2015MNRAS.446..120D,2015A&A...581A..10C,2016MNRAS.456.1359F},
 we have come to understand that in ground-based data taken with natural seeing, the effect of scattered light creates a non-uniform light screen that affects the entire image. The intensity of this light screen depends on the proximity of the emitting sources, but even in regions of the sky with few stars, virtually all pixels in the image are affected by this scattered light at a surface brightness of $\mu_V$$\sim$29 mag/arcsec$^2$ \citep[see e.g.][]{slater2009,trujillo2016,Montes2021}. For this reason, it is absolutely necessary to eliminate scattered light by constructing point spread function (PSF) models of the images where the PSF extends as far in radius as possible. 

In addition to the light screen caused by scattered light, the characterization of the extended PSF of deep images is needed to solve a number of problems in low surface brightness studies. Some examples are the detection of objects with surface brightness comparable to the intensity of the scattered light \citep[see e.g.][]{2021A&A...645A.107H,2021A&A...656A..44R,2023A&A...679A.157R,2024A&A...681A..15M}, the colour estimation of low surface brightness sources \citep{roman2020,2024A&A...684A..99G}, or the analysis of the outer part of galaxies \citep[see e.g.][]{2013MNRAS.431.1121T,sandin2014,2014MNRAS.441.2809M,2019A&A...621A.133B,2019MNRAS.483..664M}. 

We present the characterisation of the extended PSFs (R$>$5 arcmin) in the Sloan g and r bands of the Large Binocular Cameras \citep[LBC; ][]{2008A&A...482..349G} of the Large Binocular Telescope (LBT). This work is part of the LIGHTS ultra-deep optical imaging survey \citep[Large Binocular Telescope Imaging of Galactic Halos and Tidal Structures; ][]{trujillo2021,Zaritsky2024}. Unlike other multi-purpose deep imaging surveys, LIGHTS focuses on nearby galaxies and their surroundings. The data were taken in dark time and for a period of less than two hours per galaxy. LIGHTS has an observing strategy plus a data reduction pipeline specifically designed to study low surface brightness features.

To date, LIGHTS has observed 28 nearby galaxies (Distance$\lesssim$20 Mpc) down to the surface brightness limits of $\mu_g$$\sim$31.3 mag/arcsec$^2$ and $\mu_r$$\sim$30.7 mag/arcsec$^2$ \citep[3$\sigma$ in equivalent areas of 10$\times$10 arcsec$^2$;][]{Zaritsky2024}. The field of view of LIGHTS typically allows exploration of central galaxies and their satellites out to a radius of about 100 kpc. In this sense, LIGHTS is well complemented by future surveys such as LSST, which will be able to survey satellites at greater distances but with lower surface brightness due to its less homogeneous observing conditions. It will also be complemented by higher spatial resolution surveys such as EUCLID, which, although not as deep as LIGHTS in terms of surface brightness, will allow the characterisation of compact sources such as globular clusters that cannot be resolved from the ground. In short, LIGHTS is currently the deepest and most extensive ground-based survey ever undertaken of nearby galaxies and is teaching us important lessons about the handling of very deep data that will be used in the next generation of sky surveys.

To construct the extended PSFs of the LBCs, we follow a similar methodology to that used by \citet{infante2020} to construct the extended PSFs of the Sloan Digital Sky Survey \citep[SDSS;][]{2000AJ....120.1579Y}. This methodology was found to be valid for characterizing the extended PSF of large-aperture telescopes such as Subaru \citep[][]{2024MNRAS.531.2517G}. That is, we use stars of different magnitudes to generate the PSF across different radial ranges. The brighter (and therefore saturated) stars are used only for the outer part, while the fainter stars are used to generate both the inner and the intermediate parts of the PSF. This paper is organized as follows.

Section \ref{data} gives a brief overview of the data used in this work. Section \ref{constructing} is devoted to the construction and assembly of the different parts that make up the extended PSFs. In section \ref{sec:results_scatterlight}, we show a practical example of how the extended PSFs are used to remove the scattered light from the LIGHTS NGC3198 field. The magnitude system used in this work is AB \citep{oke1983}. 


\section{Data}
\label{data}

The data used to construct the extended PSFs of the LBCs are from the LIGHTS survey \citep{trujillo2021}. The LBCs are two prime focus cameras. LBC Blue is blue optimised (3500 to 6500 \text{\AA}), while LBC Red is red optimised (5500 \text{\AA} to 1$\mu$). The LBCs are used simultaneously (binocular mode). The LBC focal plane consists of four CCDs (2048 $\times$ 4608 pixels) with a pixel scale of about 0.224\arcsec. Each CCD covers about 7.8 arcmin $\times$ 17.6 arcmin, with gaps between the chips of $\sim$ 18 arcsec. The field of view is about 23 arcmin $\times$ 25 arcmin.

LIGHTS is currently a survey of 28 nearby galaxies, covering a total area of about 7 square degrees \citep[see e.g.][for a description of the first 25 galaxies analysed]{Zaritsky2024}. Each galaxy is observed simultaneously in the g and r bands for a total on-source time of $\sim$1.5\,h. The zero point of both filters has been set to 22.5 mag. The average seeing of the images is $\sim$1\arcsec. LIGHTS fields are constructed using a dithering pattern with offsets of approximately the size of the central galaxy being observed. The total field covered is typically about 40$\times$40 arcmin, although the zone of maximum depth (with 70\% of the time on the source) is about 20$\times$20 arcmin. In order to minimise the destruction of low surface brightness structures, the background subtraction in each frame is performed in each detector by removing a single constant. The limiting magnitudes for point-like sources, computed within a circular aperture with a radius equal to the Full Width Half Maximum (FWHM) of an unresolved source, in the g and r bands are $\sim$27.5 and $\sim$26.6 mag (5$\sigma$), respectively. Stars brighter than magnitude 19 are saturated in both filters (the on-source exposure time of the individual frames that make up the LIGHTS mosaics is 3 min).  Therefore, such bright stars are only used to construct the intermediate and outer parts of the extended PSFs. 

Due to the nature of the survey, the presence of bright stars is avoided as much as possible in the selection of the observed fields \citep[see ][ for further details]{Zaritsky2024}. For this reason, the number of bright stars whose characterisation allows us to construct PSFs that are as extended as possible (i.e. with radii $\sim$5 arcmin comparable to the diameter of the studied galaxies given by the location of the isophote $\mu_V$=26 mag/arcsec$^2$) is limited. In the following sections, we explain the procedure used to construct such extended PSFs.

\section{Constructing the extended LBC PSFs}
\label{constructing}

Due to the observing strategy of the LIGHTS survey, it is not possible to construct an extended PSF with a single star.  As mentioned above, stars brighter than magnitude 19 are saturated (in both filters). Therefore, if we want to characterize the full radial distribution of the PSF from the innermost arcseconds out to arcminute scales, we need to use stars with a wide range of magnitudes. To this end, we have used up to 4 different magnitude ranges, with the aim of obtaining the most accurate characterisation of the entire PSF in terms of signal-to-noise ratio. In the following, the magnitude we quote corresponds to Gaia's G-band filter. We have used the values quoted in the eDR3 release \citep{2021A&A...649A...1G}. G-band was chosen for two reasons. The first is that we need to refer to unsaturated magnitudes, and for our brighter stars, the g and r band values will not be reliable due to their saturation. The other reason is the simplicity of the procedure. The G-band is broad enough to cover both the g and r Sloan filters, so the magnitudes given in this filter are representative of the value of these stars in both Sloan filters.

The central (core) regions of the global PSFs are made up of unsaturated stars with magnitudes between 20 and 21 mag. This range is useful for describing the PSF within a radius of $\sim$2\arcsec. From 2\arcsec\ to 16\arcsec, a part we have called the intermediate region of the PSF, we use three different magnitude ranges: 17 to 18 mag ($\sim$2\arcsec\ to $\sim$7\arcsec), 13 to 15 mag ($\sim$7\arcsec\ to $\sim$13\arcsec) and 9 to 12 mag ($\sim$13\arcsec\ to $\sim$16\arcsec). Finally, the outermost parts of the PSFs are built using one of our brightest stars (7 mag), giving us radial coverage from about $\sim$16\arcsec\ to $\sim$390\arcsec. For distances greater than this, we have to extrapolate the behaviour of the PSFs. The following sections describe in detail how we build the different parts of the PSFs and how we merge them into a common and global PSF.

\subsection{Constructing the outer (16\arcsec\ to 390\arcsec)\ part of the PSFs}

As mentioned above, the selection of LIGHTS fields \citep[][]{Zaritsky2024} avoided as much as possible the proximity to bright stars. For this reason, it is difficult to find a sufficiently bright and isolated star that both avoids the central galaxy and is as far as possible from the edge of the image. After visually inspecting all LIGHTS fields, we found a suitable candidate with which to determine the outer part of PSF. The star is HD 134023 (R.A.=15:05:23.7; Dec=+55:40:38.3; J2000; G-band=7.6 mag) and is located in the field of the galaxy NGC5866 (see panel (a) in Fig.~\ref{outer1}). Fortuitously, this field has been observed during two observing campaigns and is one of our deepest fields with a total time on the source of 3.1\,h and limiting surface brightnesses of $\mu_g$$\sim$32 mag/arcsec$^2$ and $\mu_r$$\sim$31.2 mag/arcsec$^2$ (3$\sigma$; 10\arcsec$\times$10\arcsec). The construction of the outermost part of the PSF is a two-step process. 

\subsubsection{First step}
\label{outerpsfstep1}

\begin{center}
    \begin{figure*}[!t]
        \centering
        \includegraphics[width=0.51\linewidth]{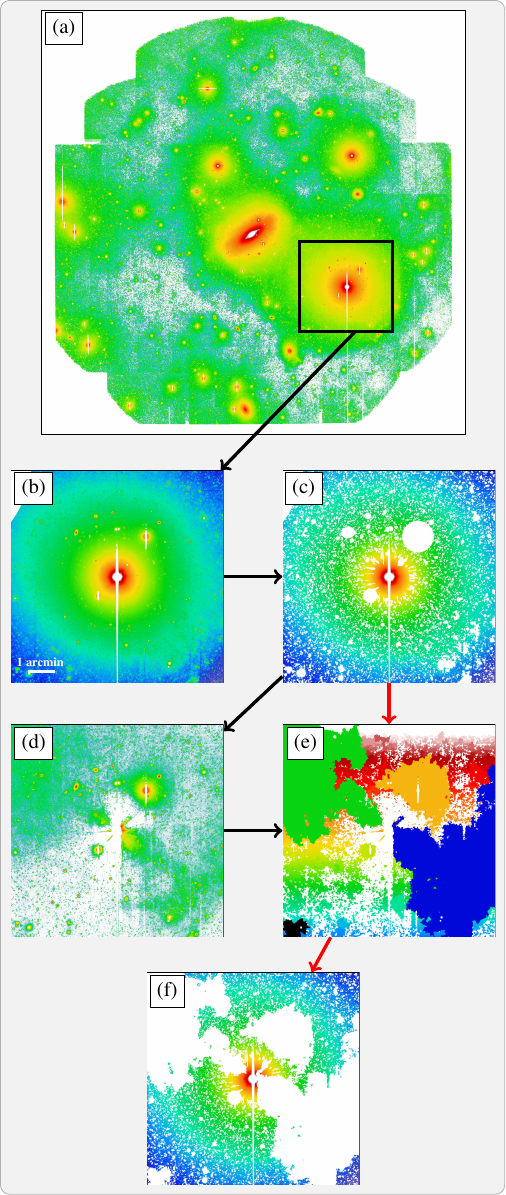}
    	\caption{ Steps taken to build the outer part of the PSF. This figure shows the g-band, but the procedure is also valid for the r-band. Panel (a) shows the position of the star HD 134023 within the LIGHTS field of the galaxy NGC5866. Panel (b) displays the area selected to create the PSF outer region. Panel (c) shows the first step of the mask. A preliminary model of the outer part of the PSF is subtracted from the real star to better characterise the contamination of the background around the star (panel (d)). In panel (e) we show the segmentation map generated by \texttt{NoiseChisel} and \texttt{Segment} applied to the subtracted image from (d). Finally, in panel (f) we show the new masked region, which is a significant improvement in terms of the ability to mask the contaminated areas shown in (d) compared to the original masked region shown in (c).}
        \label{outer1}
    \end{figure*}
\end{center}

We start by cutting an area of 9$\times$9 arcmin centred on HD 134023 (see panel (b) in Fig.~\ref{outer1}). While the scattered light from the PSF of this star extends beyond this boundary, the size of the area is chosen to minimise the effect of the light from the central galaxy (NGC5866) and the edge effects of the image. In the selected area, the star is the brightest object, but there are other fainter sources that need to be masked out to get a more accurate representation of the PSF. To build the masks of the contaminating objects we use \texttt{NoiseChisel} followed by \texttt{Segment} \citep[]{akhlaghi2015, akhlaghi2019}. In addition to this automatic mask generation, to be conservative, we augment the masks for the brightest contaminants. Three of these correspond to the three brightest targets around HD 134023 (two stars and one galaxy). These masks are shown as white circles in panel (c) of Fig.~\ref{outer1}. The fourth corresponds to the central galaxy (NGC5866). This mask appears in panel (c) as the white area in the upper left corner.  

\begin{center}
    \begin{figure*}[!t]
        \centering
        \includegraphics[width=\linewidth]{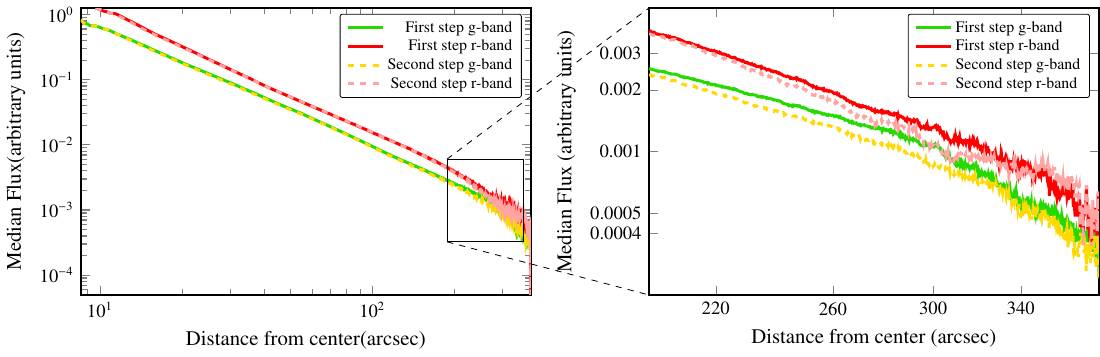}
    	\caption{ Radial profiles of the outer region of the PSF of the LBC camera in the g- and r-band filters. The left panel shows how the shape of the radial profile changes after different masking steps. The right panel zooms in on the outermost region to better visualise the effect of masking contamination from nearby sources.}
        \label{outer1-rad}
    \end{figure*}
\end{center}

Exploiting the near circular symmetry (see a detailed discussion in the Appendix) of the LIGHTS PSF, we obtain the radial profile 
of the PSF using the non-masked pixels employing \textit{astscript-radial-profile} \citep{2024RNAAS...8...22I}. Using this first version of the radial profile (see Fig. \ref{outer1-rad}; left panel), we build a preliminary circular 2D PSF model, also with the help of Gnuastro tools. This model is flux normalised in the radial interval from 22.4\arcsec\ to 44.8\arcsec\ (i.e. 100 to 200 pixels). This range is chosen to avoid the saturated part of the star and to minimise contamination from background sources. Once the model was created, we subtracted it from the real star. The result is shown in panel (d) of Fig. \ref{outer1}. 

Subtracting the first version of the model from the outer part of the PSF reveals a large number of structures that may affect the properties of the outer surface brightness profile of the PSF. We therefore re-run Gnuastro's tools on this residual image, shown in panel (d),  to mask this contamination more effectively. The detection of the low surface brightness structures that affect the determination of the PSF's external shape is shown in panel (e). The resulting new masks are shown in panel (f). 

\subsubsection{Second step}

The improved masking, shown in panel (f) of Fig. \ref{outer1}, allows us to construct a new radial profile that is significantly less affected by low surface brightness contamination around the bright star. The new profiles are shown in Fig. \ref{outer1-rad}. Note that the new profiles generally have fewer counts at a given radial distance than the original ones. This is due to the improved masking, which removes contamination from other sources. The effect is particularly significant in the outermost region of the PSF (right panel of Fig. \ref{outer1-rad}). In these regions, the effect of the contamination was particularly strong due to the proximity of the galaxy NGC5866 and other objects. 

Bleeding, due to saturation of the bright star, also affects the outer part of the recovered PSF. In Fig. \ref{outer1} the regions affected by bleeding can be seen as vertical spikes starting from the centre of the PSF. These regions are also masked when the outer parts of the PSFs are created. The procedure followed so far allows us to construct the region of the PSF that extends from about 16\arcsec\ to 390\arcsec. In the following, we will explain the generation of the PSF in the inner (R$<$16\arcsec) parts.

\subsection{Constructing the intermediate part (2\arcsec\ to 16\arcsec)\ of the PSFs}

The construction of the intermediate part of the PSF is divided into three different radial ranges. The selection of the different radial ranges is a compromise between the radial extension of the saturated core of the stars, which scales with their magnitude, and a similar signal-to-noise when averaging the profiles at different radial locations.

\subsubsection{Outer intermediate part (13\arcsec\ to 16\arcsec).}

To generate this region of the PSF, we used stars in three fields (NGC3941, NGC5248, and NGC6015) that have relatively large regions of the field of view free from contamination by very bright sources. We select stars with G-band magnitude between 9 and 12 mag and far (R$>$36\arcsec) from other bright sources of equal or greater magnitude. This "isolation" criterion ensures that contamination of the PSF is not an issue in the range of interest (i.e. 13\arcsec\ to 16\arcsec).

Using the above criteria, seven stars were selected for the outer intermediate region of the PSF. Their positions and magnitudes are given in the Table \ref{star_list}. The process of stacking them to obtain a common PSF is as follows. First, we make an image of 34\arcsec$\times$34\arcsec\  centred on each star. We normalise the radial fluxes among stars using the values from 12\arcsec\ to 16\arcsec. Once the PSFs are normalised, we combined the 2D images of the stars using the \textit{sigclip-mean} operator in GNU Astronomy Utilities. This operator applies a sigma-clipped mean filter to a dataset, removing outliers (in our case 3$\sigma$) from a set of measurements.\\

\begin{center}
    \begin{figure}[t]
        \centering
        \includegraphics[width=\linewidth]{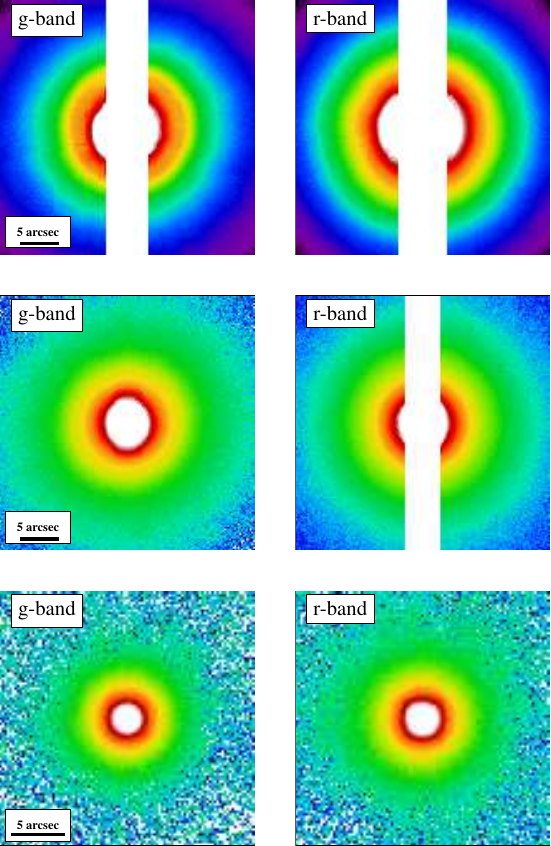}
    	\caption{The stacked images of the intermediate parts of the PSFs.
    The stacked images are in the following order. First row: outer intermediate region (13\arcsec\ to 16\arcsec). Second row: middle intermediate region (7\arcsec\ to 13\arcsec). Third row: inner intermediate region (2\arcsec\ to 7\arcsec). Each stamp has its own colour range to make it easier to see the different parts of the PSF.}
        \label{stack-int1}
    \end{figure}
\end{center}

We show the result of stacking this part of the PSF in the first row of Fig. \ref{stack-int1}. The vertical white areas correspond to regions that are masked (during the stacking process) due to the presence of bleeding (associated with the saturation of the stars), as happened with the outermost part of the PSF shown in the previous section. \\

\begingroup 
    \setlength{\tabcolsep}{10pt} 
    \renewcommand{\arraystretch}{1.5} 
    \begin{table}
        \centering
        \begin{tabular}{p{0.6cm} c c c c}
        Num & Field & RA & Dec & G (mag)  \\
       \hline \hline 
        1 & NGC5248 & 204.50033 & 9.06079 & 10.83 \\
        2 & NGC5248 & 204.57361 & 8.78941 & 10.98 \\
        3 & NGC5248 & 204.30708 & 8.73974 & 11.25 \\
        4 & NGC5248 & 204.58561 & 9.05267 & 11.30 \\
        5 & NGC5248 & 204.48749 & 8.76922 & 12.04 \\
        6 & NGC3941 & 178.41789 & 37.04541 & 10.79 \\
        7 & NGC6015 & 238.14633 & 62.31770 & 9.19 \\
       \hline \hline 
       \end{tabular}
       \caption{The list of selected stars for constructing the outer intermediate part.}
        \label{star_list}
    \end{table}
\endgroup

\subsubsection{Middle intermediate part (7\arcsec\ to 13\arcsec)}

The process of constructing this part of the PSF follows the same procedure as for the outer intermediate region, but using 63 stars in the magnitude range of 13 to 15 (G-band). We used the 18 fields of the 25 current LIGHTS fields that were fully reduced at the time of this work.  The normalisation annulus used to bring all the stars to the same flux was from 9\arcsec\ to 11\arcsec. The result of stacking the images of these stars is shown in the second row of Fig. \ref{stack-int1}.

\subsubsection{Inner intermediate part (2\arcsec\ to 7\arcsec)}

\begin{center}
    \begin{figure*}[t]
        \centering
        \includegraphics[width=\linewidth]{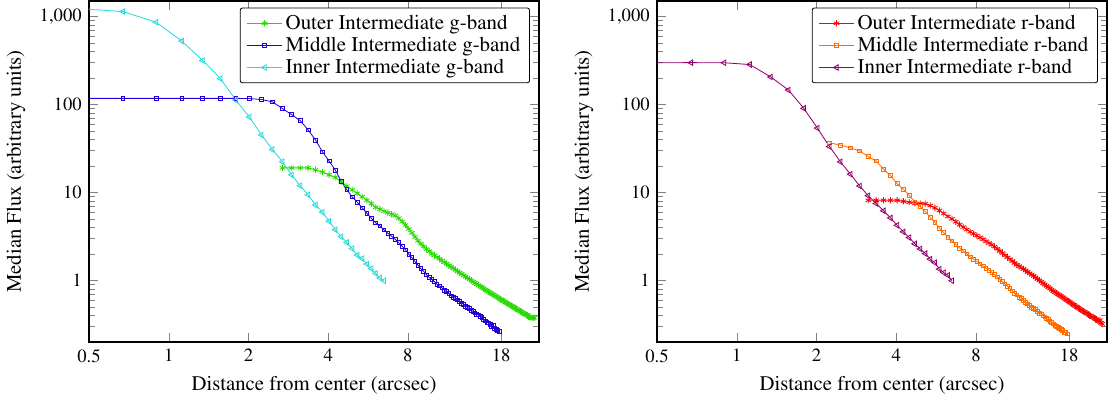}
    	\caption{The radial profiles of the intermediate parts of the PSFs in the g (left panel) and r (right panel) Sloan bands. The different fluxes are indicative of the different magnitude ranges used to construct these different radial parts of the PSFs.}
        \label{rad-int1}
    \end{figure*}
\end{center}

For this part of the PSF, we used 76 stars in the magnitude range of 17 to 18 mag (G-band). As before, we used 18 of the 25 LIGHTS fields. Because these are fainter stars, we do not lack candidate sources but need to avoid contamination of bona-fide point-like sources (i.e. stars) produced by compact background objects, such as high redshift galaxies. We add an additional criterion and require Gaia sources in this magnitude range to have a parallax measurement that is at least three times larger than the associated uncertainty. Moreover, we request the axis ratio to be at least 0.9. We use the radial range from 4.5\arcsec\  to 9\arcsec\ to normalise the profiles. The result of stacking the images of these stars is shown in the bottom row of Fig. \ref{stack-int1}. In Fig. \ref{rad-int1}, we show the radial profiles in both g and r Sloan filters of the different radial ranges of the PSFs. 

\subsection{Constructing the core  (0\arcsec\ to 2\arcsec) of the PSFs}

To create the innermost part of the PSFs, we followed the previous steps using stars with G-band magnitudes between 20 and 21. Stars in this magnitude range do not show saturation in either our g and r filters. We are more conservative in selecting stars rather than background galaxies. We add the additional requirement that the axis ratio be greater than 0.93.

We take an additional measure to remove binary stars. We exclude any star that belongs to the Washington Double Star (WDS) catalogue \citep{2012yCat....102026M}, which contains a list of 141000 known binary stars. To be even more conservative, we remove any star that has another object within 90\arcsec\ that is brighter than 25 mag in the LIGHTS catalogue (i.e. in the relevant g-band or r-band). We are left with a total of 92 and 95 stars in the g and r bands, respectively, to combine to form the core. \\

\begin{center}
    \begin{figure}[t]
        \centering
        \includegraphics[width=\linewidth]{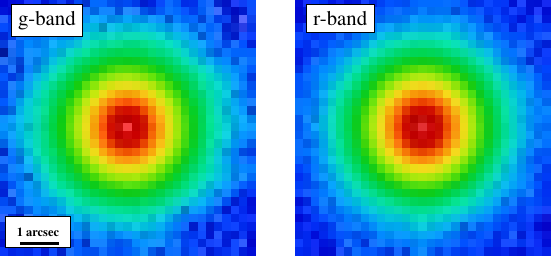}
    	\caption{The stacked images of the core part of the LBC PSFs in the g and r Sloan filters.}
        \label{core-av}
    \end{figure}
\end{center}

\begin{center}
    \begin{figure}[t]
        \centering
        \includegraphics[width=\linewidth]{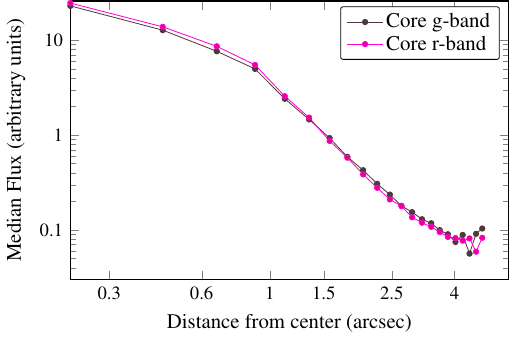}
    	\caption{The radial profiles of the core of the LBC PSFs in the g and r Sloan bands.}
        \label{rad-core-av}
    \end{figure}
\end{center}

To recover the core of the PSF, we mask all the contaminant sources within a radial range of 7\arcsec\ to our selected stars. We normalise the fluxes of each of our stars using the values within 1\arcsec\ to 2\arcsec\ prior to recovering the stacked radial profile. The final stacked images representing the core of the PSFs are shown in Fig. \ref{core-av} and their radial profiles in Fig. \ref{rad-core-av}. 

\subsection{Building the global PSFs}

\begin{center}
    \begin{figure}[t]
        \centering
        \includegraphics[width=\linewidth]{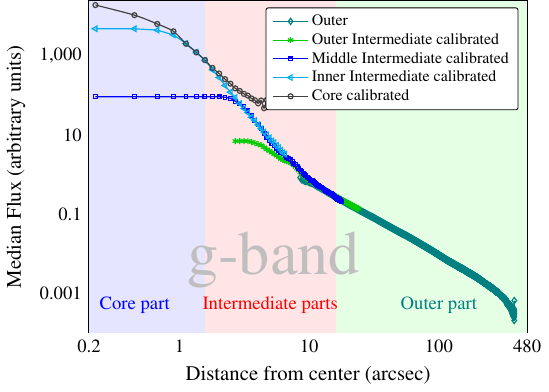}
    	\caption{Building the LIGHTS global g-band PSF. The profiles of the different parts of the PSF are matched in different regions.}
        \label{rad-un-g}
    \end{figure}
\end{center}

\begin{center}
    \begin{figure}[t]
        \centering
        \includegraphics[width=\linewidth]{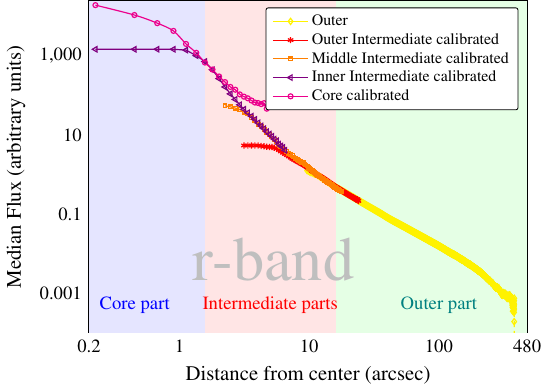}
    	\caption{Building the LIGHTS global r-band PSF. The profiles of the different parts of the PSF are matched in different regions to allow a smooth transition.}
        \label{rad-un-r}
    \end{figure}
\end{center}

\begin{center}
    \begin{figure}[t]
        \centering
        \includegraphics[width=\linewidth]{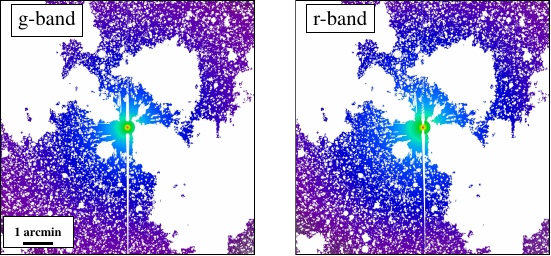}
    	\caption{The 2D image of the LIGHTS PSFs in the g- and r-bands. The white areas correspond to masked areas due to contamination from nearby sources or bleeding.}
        \label{un-img}
    \end{figure}
\end{center}

Once the process of constructing the five parts of the PSF is complete, we produce a single PSF resulting from the assembly of the different regions. We select a radial range where the outer intermediate and outer PSFs have comparable S/N (i.e. 15.7\arcsec\ to 19\arcsec). This matching is necessary to ensure a smooth transition between the different parts of the PSF. Once the outer intermediate and outer parts of the PSF have been joined, the process is repeated with the remaining parts of the PSF until the whole PSF is complete.  Fig.~\ref{rad-un-g} and Fig.~\ref{rad-un-r} show the radial regions (11.2\arcsec\ to 15.7\arcsec, 4.9\arcsec\ to 7.2\arcsec\ and 1.3\arcsec\ to 2\arcsec) where the different parts of the PSFs have been joined. The images including the different parts of the PSFs are shown in  Fig.~\ref{un-img}. The white areas correspond to the regions that have been masked due to the presence of nearby contaminants, as explained above (see section \ref{outerpsfstep1}). Fortunately, the optical path of the LBT telescope has few elements that disturb the symmetry of the PSFs. Therefore, they do not deviate significantly from circular symmetry (see the Appendix for further details). For this reason, it is possible to use the radial profile of the PSF to build a complete two-dimensional PSF model, filling in the gaps due to masking. In the next section, we will explain the circularisation process and how to extend the PSF radially.

\subsection{Creating the extended PSFs}

To fill the empty regions where we lack information due to the presence of nearby sources, we take advantage of the almost circular symmetry of the LIGHTS PSFs (see the Appendix for a discussion on the symmetry of the PSF). This allows us to use the radial profile of the PSFs and fill the empty regions with the average flux at a given radial distance.

In this work, the radial extent of the measured PSFs is ultimately limited by the brightness of the brightest star used to construct the outer part of the profile. Considering the depth of our survey in this region, this allows us to obtain profiles for the PSFs that extend up to 390\arcsec. Although measuring the PSF to this extent may seem adequate, it is insufficient for some of the galaxies. Following the criterion suggested by \citet{sandin2014},  one would like to have PSF models that are at least 2 times larger than the extension of the object of interest. For this reason, and also to be able to remove all the scattered light in our field of view, which is about 30 arcmin on each side, we decided to extend the outermost part of the PSF radially, following the slope of the outer part of the observed PSF. This extension of the PSFs beyond a radial distance of 6.5 arcmin is undoubtedly the most speculative part of our work.  In the following, we justify our choice of procedure for such an extension.

To extend the radial profile of the PSFs beyond 6.5 arcmin, we assume that the  PSFs follow the same slope as that observed in the outer region (R$>$50\arcsec) of the current PSF models. This assumption is motivated by the outer shape of the large collection of PSFs given in Fig. 1 by \citet{sandin2014}.  \citet{sandin2014} shows that in the "aureole" region\footnote{\citet{sandin2014} describes the aureole region as the radial range of the PSF between 30\arcsec\ to 1.5 degrees.}, the slope of the PSFs for different telescopes and filters is well approximated by a power-law behaviour. We fit a power-law to the PSFs over the radial range from 110\arcsec\ to 225\arcsec. This range is a compromise between having a high signal to noise of the outer part of the PSF profile and potential problems of over-subtraction of the background affecting the outermost regions of the PSFs.
\begin{center}
    \begin{figure}[t]
        \centering
        \includegraphics[width=\linewidth]{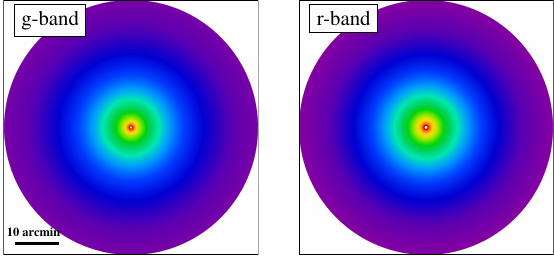}
    	\caption{Circularised and extended to a radial range of 30 arcmin of the g- and r-band LIGHTS PSFs.}
        \label{ext-img}
    \end{figure}
\end{center}

\begin{center}
    \begin{figure*}[t]
        \centering
        \includegraphics[width=\linewidth]{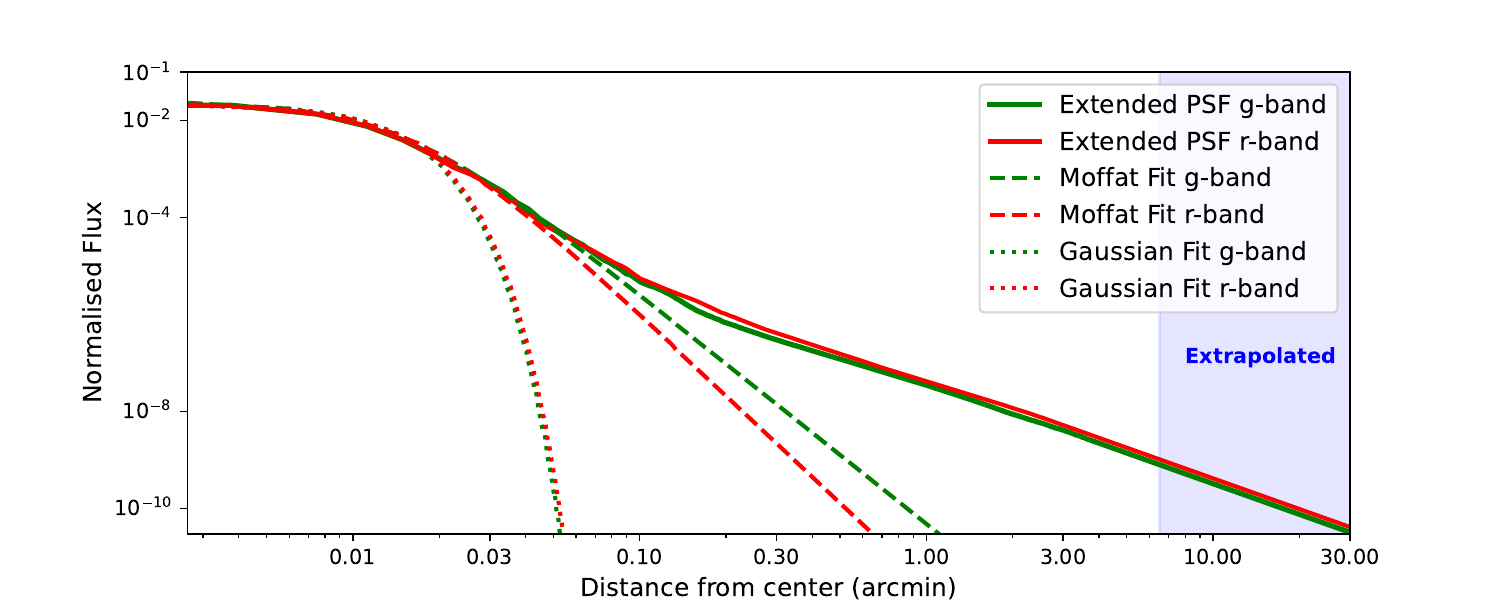}
    	\caption{Circularised and extended radial profiles of the g- and r-band LIGHTS PSFs. The region measured from stars directly in the image corresponds to R$<$6.5 arcmin. Beyond this radius, the PSFs have been extrapolated assuming a power-law decay with a slope of -2.1 for the g- and r-bands. We include both the Gaussian (FWHM$_g$=1.15$\pm$0.05\arcsec\ and FWHM$_r$=1.18$\pm$0.05\arcsec) and the Moffat (FWHM$_g$=1.00$\pm$0.05\arcsec, $\beta_g$=2.4$\pm$0.1 and FWHM$_r$=1.04$\pm$0.05\arcsec, $\beta_r$=2.8$\pm$0.1) profiles after fitting these functions to the global PSF models. These fits show where traditional approaches to modelling the PSFs \citep[see e.g.][]{2001MNRAS.321..269T,2001MNRAS.328..977T} start to deviate ($\sim$1.5\arcsec\ for Gaussian and $\sim$3\arcsec\ for Moffat fits) from the extended PSFs.}
        \label{rad-ext}
    \end{figure*}
\end{center}

For the g-band, we found a slope of -2.1$\pm$0.1, while for the r-band, the slope is -2.1$\pm$0.2. The uncertainties are estimated based on the variation in the recovered slope when making minor changes to the radial fitting range. Values for the outer PSF slope from ground-based images between -2 and -3 are common in the literature. For example, \citet{infante2020} found a slope of -2.5 for SDSS PSFs. \citet{trujillo2016} got -2.1 (r-band), while \citet{Mancera2024} got -2.9 (g-band) and -2.55 (r-band) for the GTC new OSIRIS camera, while for Gemini telescopes \citet{2024A&A...684A..99G} got -2.6 for g-band and -2.75 for r-band. We caution that the measured slope of the outer PSFs obtained using images dedicated to the study of galaxies that fill the central part of the image are necessarily steeper than the real slope of the PSF. This is due to the pipeline processing of the image, which, although designed to minimise the effect of background over-subtraction, is necessarily affected by scattered light. Finally, assuming the slopes we have measured above for the LBT telescope, we extend the PSFs up to 30 arcmin.  These circularised and extended profiles are shown in Fig.~\ref{ext-img}. The radial profiles are provided in Fig. \ref{rad-ext}.

\section{Using the global PSF to remove  scattered light in the NGC3198 field}
\label{sec:results_scatterlight}
\begin{center}
    \begin{figure}[h]
        \centering
        \includegraphics[width=0.83\linewidth]{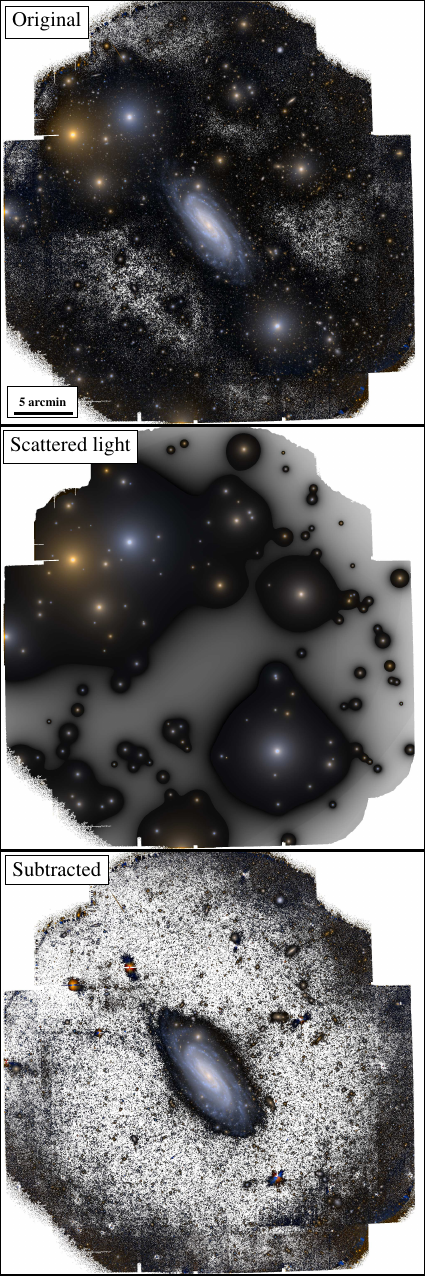}
    	\caption{Colour composite image of the LIGHTS field NGC3198, using \textsl{g} and \textsl{r} Sloan filters. The figure has been created using \texttt{astscript-color-faint-gray} \citep[][]{2024RNAAS...8...10I}. Upper panel: original image. Note the enormous contamination caused by the scattered light from the brightest stars at the edge of the central galaxy. Middle panel: modelling of the scattered light field of all stars brighter than 17 mag (G-band). Bottom panel: the field of LGHTS NGC3198 after removing the scattered light field. The outermost parts of the central galaxy are now better defined, allowing more accurate analysis of this region.}
        \label{color-img}
    \end{figure}
\end{center}

We perform the practical exercise of removing the scattered light from the LIGHTS NGC3198 field using our extended PSFs. This field is particularly attractive for this exercise because the central galaxy is affected by scattered light from several nearby stars at its edges (see Fig. \ref{color-img}).

To construct the scattered light image in the field, we model the contribution from all stars in the field with magnitude $\text{G-band}<17$ mag. The choice of magnitude is somewhat arbitrary and can be extended if necessary. We also do not model bright stars outside the field of view of this image. While this is possible, a visual inspection of the area around this galaxy using SDSS data beyond the area covered by LIGHTS shows no bright stars affecting the outermost part of the field. Therefore, for the purposes of this exercise, we consider this to be sufficient. A total of 133 stars are then modelled. The positions of these stars (all saturated) in the images are taken from the Gaia catalogues. This works because the Gaia catalogues and the LIGHTS images were taken only a few years apart.

The modelling of the scattered light field followed a hierarchical approach, starting with the PSF fitting and subtraction of the brightest star. This was then repeated for the second brightest star, and so on. This iterative procedure was necessary because the light emitted by very bright stars significantly affects the fitting and subtraction of nearby, fainter stars. The PSF fitting of each star involved two steps: the determination of the centre of the star and the flux calibration of the PSF model to the light profile of the star. The first is solved using the coordinates from the Gaia catalogue. To calibrate the PSF model to the brightness of each star, we use the radial surface brightness profile of each star. Since the brightness varies, we cannot use a fixed radial range to perform the calibration for all stars, as it could be affected (depending on the star's magnitude) by the saturation region or a poor signal-to-noise ratio. For this reason, the selected range of the star's profile is given by the surface brightness range between 22.5 and 25 mag/arcsec$^2$. This range is common for both g and r bands. Given the properties of the LIGHTS survey, this surface brightness range skips both the saturation brightness ($\sim$21.3 mag/arcsec$^2$  in g and $\sim$20.3 mag/arcsec$^2$  in r-band) and the limiting surface brightness (31.3 mag/arcsec$^2$ in g and 30.7 mag/arcsec$^2$ in r-band). Once the radial range of each individual star is selected based on the common surface brightness range, we calculate the ratio F between the counts of the surface brightness profiles of the real stars and the PSF models. The ratio F is calculated using a 3$\sigma$ clipped mean to avoid the effect of potential outliers. With the star's central position and $\text{F}$ determined, each star was fully modelled and then subtracted from the original image. This process was repeated for each star until the final faintest star considered for the scattered light map of the image was modelled and subtracted.

\begin{center}
    \begin{figure}[h]
        \centering
        \includegraphics[width=0.85\linewidth]{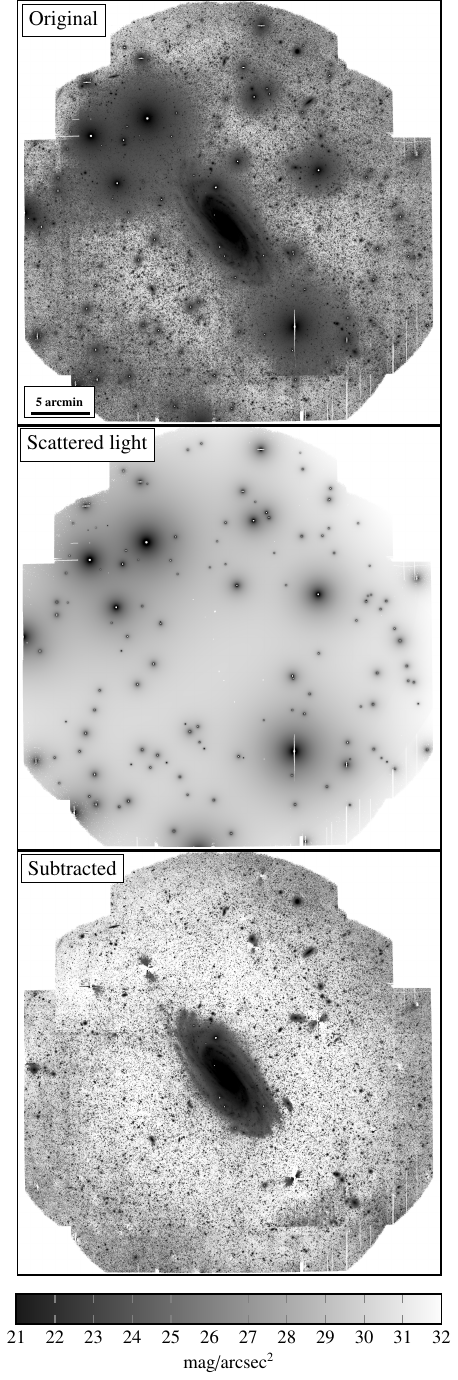}
    	\caption{Similar to Fig. \ref{color-img} but for the g-band only. The scattered light map in the middle panel shows that the contribution of the scattered light from the brightest stars in the field to the central region of the image is 28.9 mag/arcsec$^2$.}
        \label{sb-img}
    \end{figure}
\end{center} 

In the middle panel of figure \ref{color-img}, we show (in colour) the scattered light field produced by the brightest stars in the field. Note that the calibration performed produces satisfactory results in modelling not only the appearance of the scattered light but also the colours of the modelled stars. The subtraction of this scattered light produces an image (see Fig. \ref{color-img} bottom panel) that is almost free of light contamination from the bright stars in the field of view of the galaxy under study. Once the scattered light has been removed, it is easier to see the degradation of the signal-to-noise ratio towards the edge of the image (darker areas), artifacts during the reduction process, and some residuals in the core of the brightest stars, since the PSF models by construction have neither saturated cores nor bleeding problems.  This scattered light-free image also gives us the opportunity to explore the edge \citep[see e.g.][]{2020MNRAS.493...87T,2022A&A...667A..87C} of NGC3198 with unprecedented accuracy.

Having the scattered light map allows us to quantitatively explore the amount of light contamination due to the PSF effect on the nearby sources at the location of the central galaxy. In Fig. \ref{sb-img} we show this for the g-band filter. In the centre of the panel, the scattered light contributes with a screen of about 28.9 mag/arcsec$^2$. In the case of the r band, the scattered light map has a similar structure and the contamination in the centre is 28.1 mag/arcsec$^2$. Interestingly, also the colour of the scattered light at the central part of the field (g-r)$_0$=0.8 mag barely changes when adding enough stars to model it. At the edge of NGC3198, due to the distribution of the bright stars, the scattered light contamination is more important and reaches surface brightness levels comparable to the surface brightness of the outer disc of the object, making its study more complex.

\begin{center}
    \begin{figure}[h]
        \centering
        \includegraphics[width=0.8\linewidth]{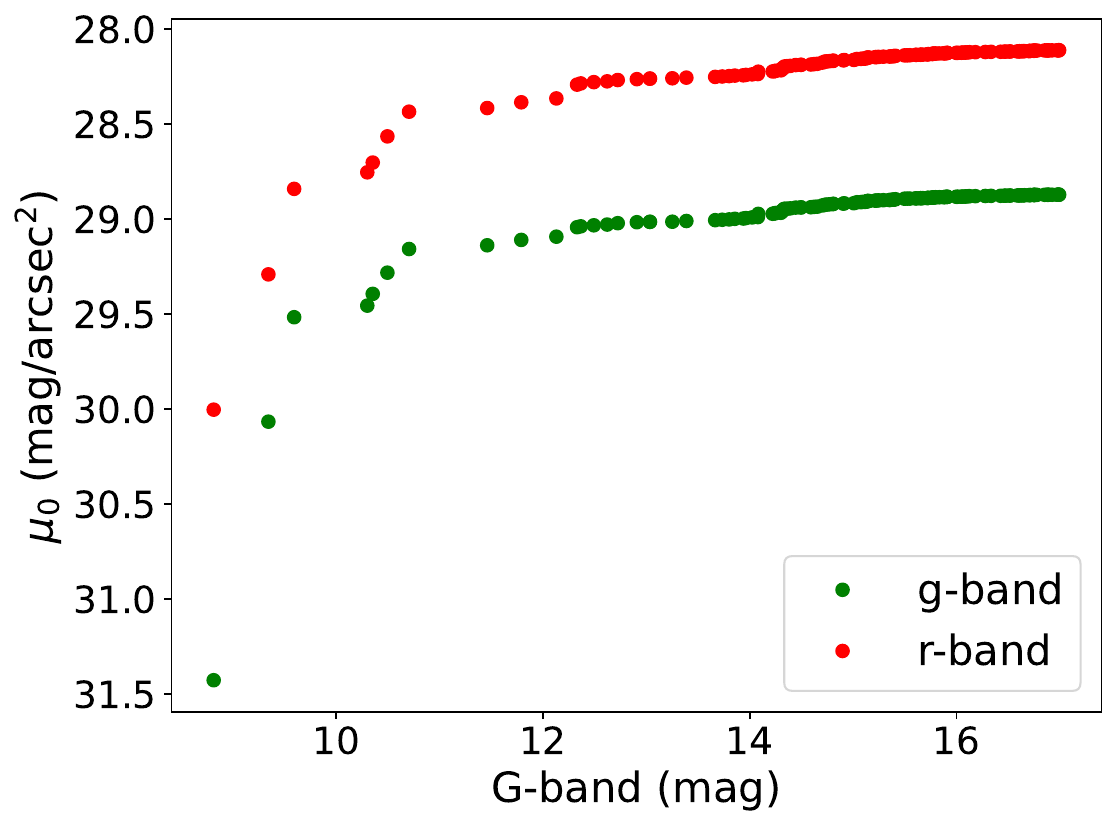}
    	\caption{The contribution of scattered light to the surface brightness at the centre of the image as the number of stars modelled increases by adding fainter stars. The surface brightness at the location of the main galaxy (here labeled as $\mu_0$) becomes brighter as fainter and fainter stars are modelled. In this particular field, modelling the scattered light of stars fainter than 16 mag (G-band) barely changes the central surface brightness of the image by less than 3\%.}
        \label{sb-img-centre}
    \end{figure}
\end{center}

A final test we can perform is to examine the contribution of the scattered light to the surface brightness in the central region of the image as a function of the magnitude of the stars being modelled. In Fig. \ref{sb-img-centre} we show how the surface brightness of the scattered light becomes brighter as fainter and fainter stars are added to model the scattered light of the field. For this particular field, the contribution to the central surface brightness remains almost unchanged ($<$3\%) when stars fainter than 16 (G-band) are added to the model. For other fields, with stars of different magnitudes and/or at different positions in the image, the shape of the curves shown in Fig. \ref{sb-img-centre} will of course be different.

\section{Conclusions}
\label{conclusions}

The new generation of ultra-deep optical sky surveys is designed to routinely obtain images with surface brightnesses of $\mu_V$$\sim$30 mag/arcsec$^2$ or fainter. One of the main problems facing the analysis of these images is the contribution of light scattered from the bright sources. As shown in the present paper, but also in previous work with ground-based data \citep[e.g.][]{slater2009,trujillo2016,Montes2021}, it is difficult for this contamination not to reach surface brightnesses of $\mu_V$$\sim$29 mag/arcsec$^2$, even in areas relatively far away from bright sources. This value is therefore brighter than the limiting surface brightness of the images and prevents full use of the data if not corrected. To overcome this problem, ultra-deep surveys must therefore characterise the extended PSF of their images with great care. In this paper, we show how this process has been carried out in the context of the LIGHTS survey with the LBT.

We have applied the methodology developed by \citet{infante2020} to the g- and r-band images of the LIGHTS survey. The characteristics of the survey allowed us to construct the PSF directly from the data up to 390\arcsec. While this extension is remarkable, it is insufficient to address our need to characterise the scattered light contribution in the full field of view of the LIGHTS mosaics ($\sim$30 arcmin on the side). For this reason, we extrapolate the behaviour of the PSFs from 6.5 arcmin to 30 arcmin. This extrapolation is justified by the behaviour of other extended PSFs taken from ground-based data. 

Finally, as a practical example, we show how to subtract the scattered light contamination from one of our LIGHTS fields. However, the use of the extended PSFs does not end there, and in future papers, we will show how the LIGHTS extended PSFs can be used to correct for example: a) the effect of the scattered light produced by the galaxies themselves on the study of the outermost regions of these objects, and b) the effect on the number counts of faint objects detected in the images (particularly those close to the brightest stars).

\section*{Acknowledgments}
We thank the referee for a detailed reading of the manuscript, which helped to clarify and improve the presentation of the results of this work. We thank Mohammad Akhlaghi and Sergio Guerra Arencibia for interesting discussions during the development of this work. IT acknowledges support from the ACIISI, Consejer\'{i}a de Econom\'{i}a, Conocimiento y Empleo del Gobierno de Canarias and the European Regional Development Fund (ERDF) under a grant with reference PROID2021010044 and from the State Research Agency (AEI-MCINN) of the Spanish Ministry of Science and Innovation under the grant PID2022-140869NB-I00 and IAC project P/302302, financed by the Ministry of Science and Innovation, through the State Budget and by the Canary Islands Department of Economy, Knowledge, and Employment, through the Regional Budget of the Autonomous Community. This research also acknowledge support from the European Union through the following grants: "UNDARK" and "Excellence in Galaxies - Twinning the IAC" of the EU Horizon Europe Widening Actions  programmes (project numbers 101159929 and 101158446). Funding for this work/research was provided by the European Union (MSCA EDUCADO, GA 101119830). Views and opinions expressed are however those of the author(s) only and do not necessarily reflect those of the European Union or European Research Executive Agency (REA). Neither the European Union nor the granting authority can be held responsible for them. RIS acknowledges financial support from the Spanish Ministry of Science and Innovation through the project PID2022-138896NA-C54. GG acknowledges support from IAC project P/302304 and through the IAC project TRACES which is partially supported through the state budget and the regional budget of the Consejeria de Economia, Industria, Comercio y Conocimiento of the Canary Islands Autonomous Community. MM acknowledges support from the project PCI2021-122072-2B, financed by MICIN/AEI/10.13039/501100011033, and the European Union “NextGenerationEU”/RTRP, and from the project RYC2022-036949-I financed by the MICIU/AEI/10.13039/501100011033 and by FSE+. 
J.R acknowledges financial support from the Spanish Ministry of Science and Innovation through the project PID2022-138896NB-C55. 
SR acknowledges the support of the grant PID2023-150393NB-I00 from the Spanish Ministry of Science, Innovation and Universities and SZ acknowledges the support by
 PID2021-123313NA-I00 of MICIN/AEI/10.13039/501100011033/FEDER, UE.\\
This research has made use of the Washington Double Star Catalog maintained at the U.S. Naval Observatory.

This research was done with the following free software programs and libraries:  1.23, Bzip2 1.0.8, CFITSIO 4.1.0, CMake 3.24.0, cURL 7.84.0, Dash 0.5.11-057cd65, Discoteq flock 0.4.0, Expat 2.4.1, File 5.42, Fontconfig 2.14.0, FreeType 2.11.0, Git 2.37.1, GNU Astronomy Utilities 0.20.27-9b95b \citep{gnuastro,akhlaghi2019}, GNU Autoconf 2.71, GNU Automake 1.16.5, GNU AWK 5.1.1, GNU Bash 5.2-rc2, GNU Binutils 2.39, GNU Bison 3.8.2, GNU Compiler Collection (GCC) 12.1.0, GNU Coreutils 9.1, GNU Diffutils 3.8, GNU Emacs 28.1, GNU Findutils 4.9.0, GNU gettext 0.21, GNU gperf 3.1, GNU Grep 3.7, GNU Gzip 1.12, GNU Integer Set Library 0.24, GNU libiconv 1.17, GNU Libtool 2.4.7, GNU libunistring 1.0, GNU M4 1.4.19, GNU Make 4.3, GNU Multiple Precision Arithmetic Library 6.2.1, GNU Multiple Precision Complex library, GNU Multiple Precision Floating-Point Reliably 4.1.0, GNU Nano 6.4, GNU NCURSES 6.3, GNU Readline 8.2-rc2, GNU Scientific Library 2.7, GNU Sed 4.8, GNU Tar 1.34, GNU Texinfo 6.8, GNU Wget 1.21.2, GNU Which 2.21, GPL Ghostscript 9.56.1, Help2man , Less 590, Libffi 3.4.2, Libgit2 1.3.0, libICE 1.0.10, Libidn 1.38, Libjpeg 9e, Libpaper 1.1.28, Libpng 1.6.37, libpthread-stubs (Xorg) 0.4, libSM 1.2.3, Libtiff 4.4.0, libXau (Xorg) 1.0.9, libxcb (Xorg) 1.15, libXdmcp (Xorg) 1.1.3, libXext 1.3.4, Libxml2 2.9.12, libXt 1.2.1, Lzip 1.23, OpenSSL 3.0.5, PatchELF 0.13, Perl 5.36.0, pkg-config 0.29.2, podlators 4.14, Python 3.10.6, util-Linux 2.38.1, util-macros (Xorg) 1.19.3, WCSLIB 7.11, X11 library 1.8, XCB-proto (Xorg) 1.15, xorgproto 2022.1, xtrans (Xorg) 1.4.0, XZ Utils 5.2.5 and Zlib 1.2.11.
The \LaTeX{} source of the paper was compiled to make the PDF using the following packages: biber 2.19, biblatex 3.19, bitset 1.3, caption 66580 (revision), courier 61719 (revision), csquotes 5.2n, datetime 2.60, ec 1.0, etoolbox 2.5k, fancyhdr 4.1, fancyvrb 4.5a, fmtcount 3.07, fontaxes 1.0e, footmisc 6.0d, fp 2.1d, helvetic 61719 (revision), hyperref 7.00v, kastrup 15878 (revision), letltxmacro 1.6, logreq 1.0, mweights 53520 (revision), natbib 8.31b, newtx 1.71, pdfescape 1.15, pdftexcmds 0.33, pgf 3.1.10, pgfplots 1.18.1, preprint 2011, setspace 6.7b, sttools 3.0, tex 3.141592653, texgyre 2.501, times 61719 (revision), titlesec 2.14, trimspaces 1.1, txfonts 15878 (revision), ulem 53365 (revision), xcolor 2.14, xkeyval 2.9 and xstring 1.85. 
We are very grateful to all their creators for freely providing this necessary infrastructure. This research  (and many other projects) would not be possible without them.


\section{Data availability}

The extended PSFs FITS files of the LIGHTS survey taken with the LBT in g and r Sloan bands are available at 
https://doi.org/10.5281/zenodo.15263712.

\bibliographystyle{mnras}
\bibliography{oja_template}

\begin{appendix}

\section{Deviations from the circularity of the LIGHTS extended PSFs}
\label{appendix}

Throughout this paper, we have assumed that the shape of the LIGHTS PSF is circularly symmetric. This assumption is based on two aspects. On the one hand, the optics of the LBT are quite simple, with only one arm supporting the secondary mirror of the telescope in addition to a monolithic mirror. In addition, the LIGHTS data are the result of combining multiple images of a dithering pattern, so it is expected that possible deviations from circularity due to different optical paths (i.e. off-axis variations of the PSF) are compensated in this combined PSF of the final LIGHTS mosaics. In this appendix, we quantify this hypothesis of model circularity by investigating how much the two-dimensional light distributions of real stars deviate from circularity.

\begin{center}
    \begin{figure*}[h]
        \centering
        \includegraphics[width=\linewidth]{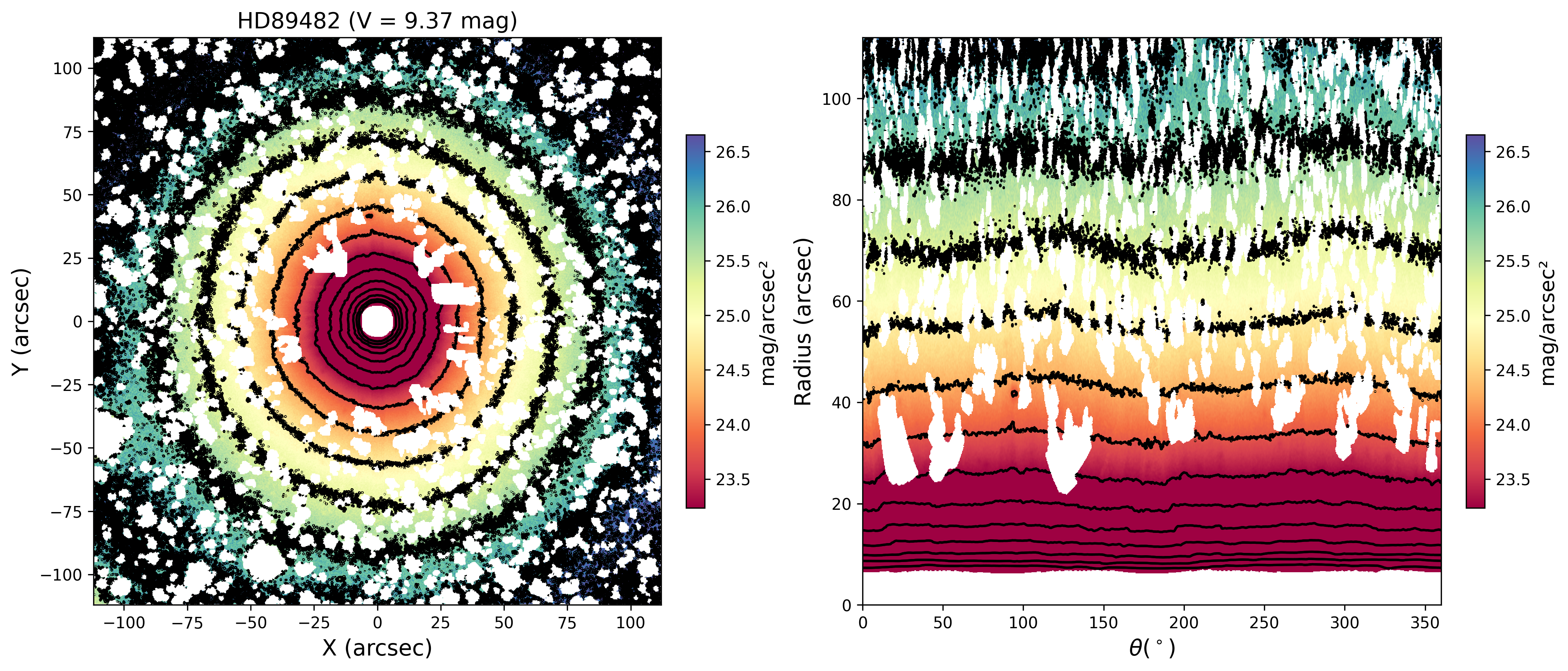}
    	\caption{Isophotal contours of the star HD89482 in both Cartesian (left panel) and polar coordinates (right panel). A perfect circular symmetry will show the isophotes as straight horizontal lines in the polar map representation. We find small deviations from circularity depending on the radial distance. The white areas correspond to regions on the maps that have been masked either by saturation in the central region of the star or by the presence of contaminant sources.}
        \label{starisophotes}
    \end{figure*}
\end{center}

The procedure we used to quantify the deviations from circularity of the real stars in our data is exemplified in Fig. \ref{starisophotes}. The star HD 89482 (V=9.37 mag) is shown in this figure. In the left panel, we show the isocontours corresponding to different surface brightnesses. The right panel shows the same image but in polar coordinates. If the light distribution of the star were perfectly circular, we would expect the isophotes in the polar map to correspond to straight horizontal lines. In practice, the isophotes deviate from the straight line by a certain radial distance. To quantify this deviation, for a given isophote value, we measure the mean radial distance of the points representing the isophote $<$R$>_{iso}$. We also measure the standard deviation of all radial distances of the points corresponding to such an isophote with respect to the mean radial distance $\sigma_{iso}$. The deviation of the circularity is characterized as $\sigma_{iso}$/$<$R$>_{iso}$.

\begin{center}
    \begin{figure*}[h]
        \centering
        \includegraphics[width=\linewidth]{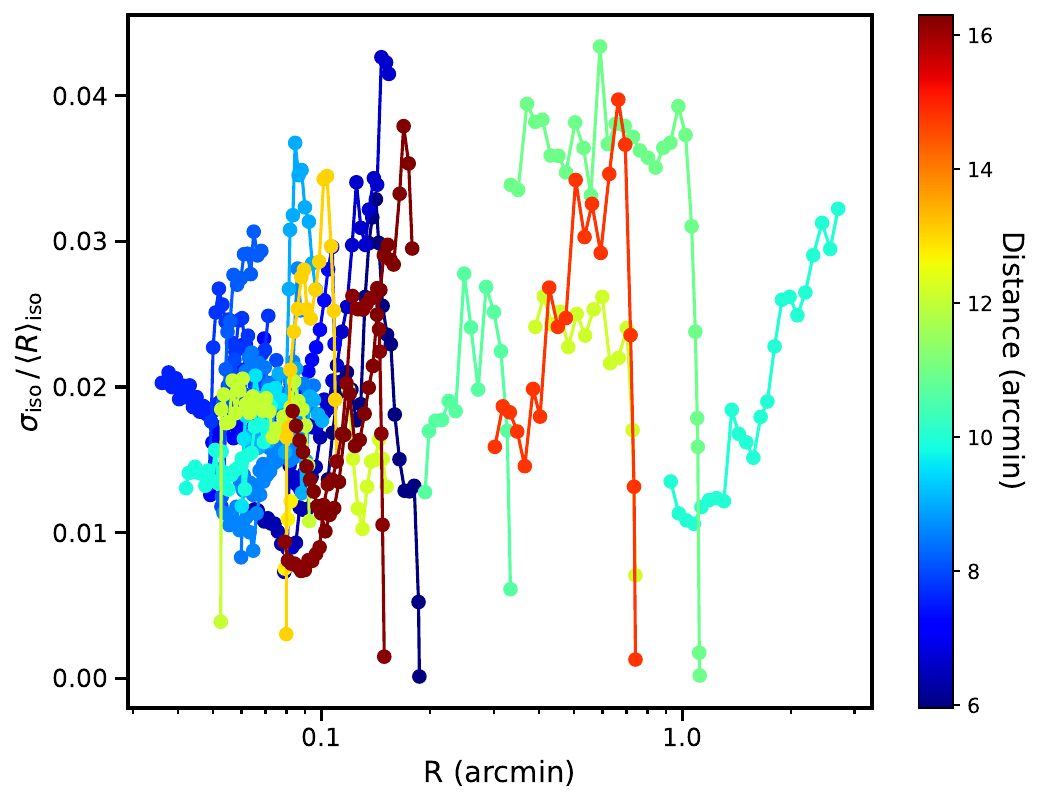}
    	\caption{The quantity $\sigma_{iso}$/$<$R$>_{iso}$, which characterizes how much the two-dimensional distributions of the real stars in the LIGHTS fields deviate from circular symmetry as a function of the radial distance of the PSF. The stars are colour-coded according to their distance from the center of the mosaic in which they are located.  We find that the deviations from circular symmetry do not depend much on the radial distance of the PSF, nor on their position in the image, and are around 2.5\%.}
        \label{circularity}
    \end{figure*}
\end{center}

In Fig. \ref{circularity} we plot for a number of real stars of different magnitudes (7.6 to 16.8 mag; G-band) how the quantity $\sigma_{iso}$/$<$R$>_{iso}$ behaves along the radial distance of the PSF. We find that typical deviations from the assumption of a perfectly circular PSF are around 2.5\%. We also find no obvious trend with respect to the distance of the star from the center of the mosaic where it is located. The radial range of the profiles shown in Fig. \ref{circularity} depends on the magnitude of the stars. Fainter (i.e. unsaturated) stars trace the innermost 0.2 arcmin of the PSF, while the outer regions of the PSF are given by the brightest stars.

\end{appendix}

\end{document}